\documentclass[prb,showpacs,twocolumn,amsmath,amssymb,floatfix]{revtex4-1}
\usepackage{graphicx}
\usepackage{bm}
\usepackage{bbm}
\usepackage{color}
\usepackage{color}
\usepackage{footnote}
\usepackage{epstopdf}
\usepackage{hyperref}
\usepackage{upgreek}
\usepackage{appendix}
\usepackage[applemac]{inputenc}

\newcommand{\comm}[1]{#1}

\newcommand{\editP}[1]{#1}
 
\begin{document}
\title{Majorana splitting from critical currents in Josephson junctions}
\author{Jorge Cayao$^{1}$, Pablo San-Jose$^2$, Annica M. 
Black-Schaffer$^{1}$, Ram\'{o}n Aguado$^2$, Elsa Prada$^3$}
\affiliation{$^1$Department of Physics and Astronomy, Uppsala University, Box 516, S-751 20 Uppsala, Sweden\\$^2$Instituto de Ciencia de Materiales de Madrid (ICMM-CSIC), Cantoblanco, 28049 Madrid, Spain\\$^3$Departamento de F\'{i}sica de la Materia Condensada, Condensed Matter Physics Center (IFIMAC) \& Instituto Nicol\'{a}s Cabrera, Universidad Aut\'{o}noma de Madrid, E-28049 Madrid, Spain}

\date{\today} 
\begin{abstract}
%The superconducting proximity effect in
\comm{A semiconducting nanowire with strong Rashba spin-orbit coupling and coupled to a superconductor can be tuned by an external Zeeman field into a topological phase with Majorana zero modes.}
Here we theoretically investigate how this exotic \comm{topological superconductor phase} manifests in Josephson junctions based on such \comm{proximitized} nanowires. In particular, we focus on critical currents in the short junction limit ($L_{\rm N}\ll \xi$, where $L_{\rm N}$ is the junction length and $\xi$ is the superconducting coherence length) and show that they contain important information about nontrivial topology and Majoranas. This includes signatures of the gap inversion at the topological transition and a \editP{unique} oscillatory pattern that originates from Majorana \editP{interference}. Interestingly, this pattern can be modified by tuning the transmission across the junction, \comm{thus providing %an alternative detection scheme 
\editP{complementary} evidence of Majoranas and their energy splittings beyond standard tunnel spectroscopy experiments}, while offering further tunability by virtue of the Josephson effect.
%remain finite with a robust feature at the topological transition and, quite remarkably, develop an oscillatory pattern in the topological phase originated from the overlaps between Majorana bound states. We also show that such oscillations are strongly affected by tuning the transmission across the junction, doubling the period of such oscillations in the tunnel regime.We also discuss the effect of finite temperature on boths supercurrents and critical currents.

\end{abstract}
\maketitle
\section{Introduction}
\label{intro}
The theoretical proposals \cite{PhysRevLett.105.177002,Lutchyn:PRL10} for artificially creating topological superconductors  \cite{kitaev} out of semiconducting nanowires with strong spin-orbit (SO) coupling and proximitized with conventional $s$-wave superconductors have spurred an unprecedented experimental effort towards realizing such exotic phase with mundane materials \footnote{For a review of the experimental state-of-the-art see Ref. \onlinecite{Aguado:RNC17}}. The topological superconductor phase is expected to occur when an external Zeeman field exceeds a critical field $B_{\rm c}\equiv\sqrt{\Delta^{2}+\mu^{2}}$, defined in terms of the induced superconducting gap $\Delta$ and the chemical potential $\mu$ of the wire. Since the electron density can be modified by external electrostatic gates and the external Zeeman field can be larger than the induced superconducting gap, the topological phase can in principle be tuned in situ. Such nontrivial phase is characterized by the emergence of non-Abelian Majorana bound states (MBSs) localized at the nanowire ends and at zero energy. The emergence of MBSs is predicted to give rise to perfect Andreev reflection, and thus perfectly quantized conductance $G=2e^2/h$ at zero bias voltage, in normal-superconductor junctions.\cite{Law:PRL09}

During the last few years, this idea has led to a vast number of important experiments in hybrid superconductor-semiconductor systems looking for such zero-bias anomaly (ZBA) in the differential conductance as the Zeeman field increases.\cite{Mourik:S12,xu,Das:NP12,Finck:PRL13,Churchill:PRB13,Lee:13} However, and despite the initial excitement, the interpretation of these experiments in terms of Majoranas has been hotly debated owing to the difficulty in ruling out alternative explanations such as parity crossings of Andreev levels,\cite{Lee:PRL12,Lee:13,Zitkoetal} Kondo 
physics,\cite{Lee:PRL12,Finck:PRL13,Churchill:PRB13,Zitkoetal} disorder,\cite{Pientka:PRL12,Bagrets:PRL12,Liu:PRL12,Rainis:PRB13,Sau:13} or smooth confinement.\cite{Prada:PRB12,Kells:PRB12} 
The nagging question of whether MBSs have been observed in nanowires and the so-called soft gap problem\cite{Mourik:S12,xu,Das:NP12,Finck:PRL13,Churchill:PRB13,Lee:13} have resulted in a second generation of  experiments, with improved devices, which show more convincing signatures of Majorana zero modes.\cite{Albrecht16,zhang16,Deng16} Despite these improvements, the topological origin of the ZBA is still under debate since subtler effects, such as the sticking of Andreev levels owing to non-homogeneous chemical potentials,\cite{StickDas17} can mimic Majoranas. Therefore, it would be very useful to study other geometries and signatures \cite{Prada:PRB17} beyond ZBAs in normal-superconductor junctions.

The direct measurement of a $4\pi$-periodic Josephson effect in nanowire-based superconductor-normal-superconductor (SNS) junctions may provide one of such strong signatures. Unfortunately, the $4\pi$ effect, which appears as a result of protected fermionic parity crossings as a function of the superconducting phase difference ($\phi$) in the junction,  is very hard to observe experimentally. First, the unavoidable presence of two additional MBSs at the wire ends (outer Majoranas), and their overlap with MBSs at the junction (inner Majoranas), renders the Josephson effect  2$\pi$-periodic.\cite{San-Jose:11a}  A second problem is stochastic quasiparticle poisoning which spoils fermion parity conservation. 
The $4\pi$-periodic Josephson effect may still be detected by out-of-equilibrium measurements, such as ac Josephson radiation or noise,\cite{Badiane:PRL11,San-Jose:11a,Pikulin:PRB12} as demonstrated recently in Josephson junctions fabricated with quantum spin Hall edges.\cite{Deacon17,Bocquillon17} Another route is offered by the multiple Andreev reflection mechanism in a voltage-biased SNS junction\cite{SanJoseNJP:13} which in current experiments\cite{Kjaergaard17,Goffman17} could provide stronger evidence of MBSs.

In this work we take an arguably easier route and focus on equilibrium supercurrents in SNS junctions which, despite the overall 2$\pi$-periodicity in the ground state, still contain useful information about the nontrivial topology and MBSs in the junction. In particular, we study short junctions with $L_{\rm N}\ll \xi$, where $L_{\rm N}$ is the normal region length and $\xi$ is the superconducting coherence length. 
Our main findings are summarized in Figs. \ref{Fig4} and \ref{Fig5}. We find that the critical current $I_{\rm c}$ through a short junction  between two finite length superconductors [two proximitized regions, each of length $L_{\rm S}$, see Fig. \ref{Fig1}(a)] and tunable normal transmission $T_{\rm N}$, shows a distinct feature at the gap inversion for the critical Zeeman field $B=B_{\rm c}$ [Fig. \ref{Fig4}(a)]. For $B<B_{\rm c}$, the critical current does not depend on the superconducting region length $L_{\rm S}$, as expected for a trivial ballistic junction in the short junction regime.\cite{Beenakker:92}  In contrast, the critical current in the topological regime increases with $L_{\rm S}$. This unique length-dependence in a short SNS junction is a direct consequence of the Majorana overlap on each S region of the junction, with one MBS at each end of S. For long enough wires ($L_{\rm S}\gg2\xi_{\rm M}$ with $\xi_{\rm M}$ being the Majorana localization length) and small transmission $T_{\rm N}\ll 1$, we obtain a re-entrant behavior of $I_{\rm c}$ as a function of Zeeman field. This reentrance results from an enhanced Majorana-mediated critical current  $I_{\rm c}\sim\sqrt{T_{\rm N}}$ (for $B>B_{\rm c}$) relative to that of conventional Andreev levels $I_{\rm c}\sim T_{\rm N}$ (for $B<B_{\rm c}$), as also discussed in Refs. [\onlinecite{PhysRevLett.112.137001, tiira17}].%As the length $L_{\rm S}$ increases, the asymptotic limit $I_{\rm c}(B\gg B_{\rm c})/I_{\rm c}(B=0)\rightarrow 1/2$ is reached which directly reflects the transition from spinful to spinless bands as the nanowire becomes topological \cite{SanJoseNJP:13}. \comm{[Why is this not apparent in Fig. 4a? This statement makes an additional and unrealistic assumption ($E_\mathrm{SO}\to\infty$)]}

On the other hand, we find that the critical current for $B>B_{\rm c}$ exhibits oscillations derived from the oscillatory energy splitting of overlapping Majoranas around zero energy. The oscillations with $B$ exhibit a surprising period doubling effect as the junction transparency is reduced  [Fig. \ref{Fig5}], which can be understood from the hybridisation properties of the four  Majoranas in the system. For small $T_{\rm N}$, the oscillatory pattern is that of two nearly decoupled finite-length topological wires with two overlapping Majoranas at the ends of each S region. As the transparency increases, the oscillations evolve into a more complex pattern which, at perfect transparency $T_{\rm N}\rightarrow 1$, results in the competing hybridization of the four Majoranas, with two energy splittings oscillating out-of phase. 
%\textcolor{red}
\comm{All these findings remain robust under finite temperatures (below the induced gap), as well as under the presence of disorder.} 
{Our results constitute a comprehensive picture, extending previous studies\cite{PhysRevLett.112.137001, tiira17}, of the effects of Majoranas on the $I_{\rm c}$ of short SNS junctions. We include the dependence of $I_{\rm c}$ with Zeeman field $B$, superconductor length $L_{\rm S}$ and transparency $T_{\rm N}$.}

We conclude that critical currents reveal signatures of Majorana splittings and, despite the finite lengths involved in the geometry of the junction, for $B>B_{\rm c}$ they contain distinct signatures of the \emph{bulk topological transition}, derived from the specific four-Majorana hybridization pattern. These findings hold relevance in view of recent proposals for topological quantum computation that crucially rely on controlling the Josephson coupling in nanowire-based Josephson junctions.\cite{Aasen16,Plugge17}
%Our findings thus constitute an interesting alternative beyond ZBAs in tunneling measurements while offering further tunability by virtue of the Josephson effect. Furthermore, they hold relevance in view of recent proposals for topological quantum computation that crucially rely on controlling the Josephson coupling in nanowire-based Josephson junctions.\cite{Aasen16,Plugge17}

The paper is organized as follows. In Sec.\,\ref{sec1} we describe the model for SNS junctions based on nanowires with SO.
Then, in Sec.\,\ref{sec13} we investigate the Zeeman field dependence of critical currents 
and discuss the emergence of oscillations as a result of MBSs overlaps. 
In Sec.\,\ref{concl} we present our conclusions. %\textcolor{red}
\comm{In Appendix \ref{Appendix} we discuss the robustness of our findings under finite temperature and disorder. For completeness, we also include a short discussion about the Majorana localization length and its typical values for the parameters used in this work.}

\section{SNS junctions made of semiconducting nanowires}
\label{sec1}
\subsection{Model}
We consider a single channel semiconducting nanowire with Rashba SO whose Hamiltonian is given by
\begin{equation}
\label{H0Hamil}
H_{0}=\frac{p^{2}_{x}}{2m}-\mu-\frac{\alpha_{\rm R}}{\hbar}\sigma_{y}p_{x}+B\sigma_{x}\,,
\end{equation}
where $p_{x}=-i\hbar\partial_{x}$ is the momentum operator, $\mu$ is the chemical potential that determines the filling of the nanowire, $\alpha_{\rm R}$ represents the strength of Rashba SO (which results in a SO energy $E_{\rm{SO}}=\frac{m\alpha^2_{\rm R}}{2\hbar^2}$), $B=g\mu_{\rm B}\mathcal{B}/2$ is the Zeeman energy resulting from the applied magnetic field $\mathcal{B}$, $g$ is the wire's $g$-factor and $\mu_{\rm B}$ the Bohr magneton.
Notice that the Zeeman and the SO axes are perpendicular to each other. 
The Hamiltonian given in Eq.\,(\ref{H0Hamil}) is discretized into a tight-binding lattice following standard methods.\cite{PhysRevB.91.024514} In all our calculations we consider typical values for InSb nanowires \cite{Mourik:S12} that include the electron's effective mass $m=0.015m_{e}$, with $m_{\rm e}$ the mass of the electron, and the SO strength $\alpha_{\rm R}=20$\,meV~nm, unless otherwise stated. 

\begin{figure}
%\begin{minipage}[t]{\linewidth}
\centering
\includegraphics[width=.49\textwidth]{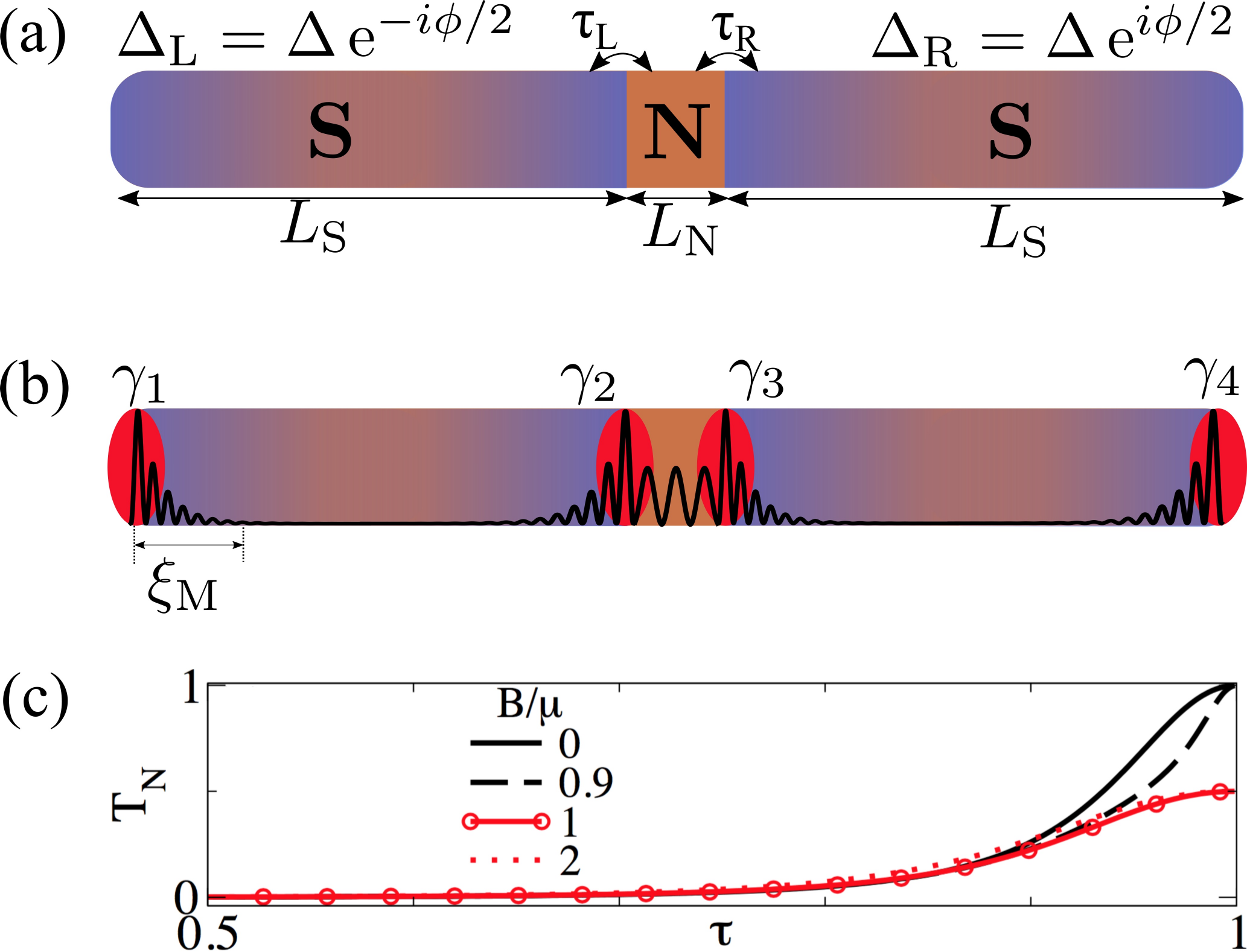} 
\caption[Sketch of an SNS junction with four MBSs]{(Color online) Sketch of a SNS junction with S and N regions of finite length $L_{\rm S}$ and $L_{\rm N}$. The transmission across the junction can be controlled by the parameters $\uptau_{\rm L(R)}\in[0,1]$, which parametrize the coupling between neighboring sites at the interfaces. 
(b) For $B>B_{\rm c}$, the S regions become topological and the system hosts four MBSs close to zero energy at $\phi=\pi$: two outer $\gamma_{1,4}$ and two inner  $\gamma_{2,3}$ ones. The MBSs are localized at the ends of the S regions and their wavefunctions exhibit an oscillatory exponential decay into the bulk of S with a localization length $\xi_{\rm M}$. (c) Relation between normal transmission $T_{\rm N}$ and coupling parameter $\uptau_{\rm L(R)}=\uptau$ for different values of the Zeeman field. Parameters: $\Delta=0$ and $\mu=0.5$\,meV, $L_{\rm N}=20$\,nm and $\alpha_{\rm R}=20$\,meVnm.} 
\label{Fig1}
%\end{minipage}
\end{figure}
%\section{SNS junctions made of semiconducting nanowires}
%\label{sec1}
In order to model the SNS junction, we now assume that the left and right regions of the nanowire are in good contact with two parent $s$-wave superconductors. Due to the proximity effect, this results in a nanowire containing left (${\rm S}_{\rm L}$) and right (${\rm S}_{\rm R}$) regions of length $L_{\rm S}$ with induced superconducting pairing potential $\Delta_{\rm S_{\rm L,R}}$, leaving the central region, of length $L_{\rm N}$, in the normal state (N). 
This model is relevant in recent experiments where hard gaps are proximity-induced into nanowires.\footnote{The quality of the semiconductor-superconductor interface in nanowire devices has been improved dramatically during the last few years.\cite{chang15,Krogstrup15,zhang16,Woerkom17} This results in a very good proximity effect with induced superconducting gaps without residual subgap conductance. Moreover, very good transparencies, down to the single-channel limit, have been reported with almost perfect Andreev conductances\cite{chang15,zhang16} $G\sim 4e^2/h$.
}

The full Hamiltonian describing the SNS junction is given in Nambu space as \cite{PhysRevB.91.024514}
\begin{equation}
\label{hsnss}
H_{\rm SNS}=
\begin{pmatrix}
h_{\rm SNS}&\Delta_{\rm SNS}\\
\Delta^{\dagger}_{\rm SNS}&-h_{\rm SNS}^{*}
\end{pmatrix}\,,
\end{equation}
where $h_{\rm SNS}$ and $\Delta_{\rm SNS}$ are matrices in the junction space and read
\begin{equation}
\label{hsnszero}
\begin{split}
h_{\rm SNS}&=
\begin{pmatrix}
H_{\rm S_{L}}&H_{\rm S_{L}N}&0\\
H^{\dagger}_{\rm S_{L}N}&H_{\rm N}&H_{\rm NS_{R}}\\
0&H_{\rm NS_{R}}^{\dagger}&H_{\rm S_{R}}
\end{pmatrix}\,,\\
\Delta_{\rm SNS}&=
\begin{pmatrix}
\Delta_{\rm S_{L}}&0&0\\
0&0&0\\
0&0&\Delta_{\rm S_{R}}
\end{pmatrix}\,.
\end{split}
\end{equation}
All the elements in the diagonal of $h_{\rm SNS}$ have the structure of $H_{0}$ given by Eq. (\ref{H0Hamil}), with chemical potential $\mu_{\rm S_{L,R}}$ for $H_{\rm S_{L,R}}$ and $\mu_{\rm N}$ for $H_{\rm N}$.
The induced pairing potentials in the left and right superconducting regions are, respectively, $\Delta_{\rm S_{L}}=\bar{\Delta}{\rm e}^{-i\phi/2}$ and $\Delta_{\rm S_{R}}=\bar{\Delta}{\rm e}^{i\phi/2}$, where $\bar{\Delta}={\rm i}\sigma_{y}\Delta\,$.  %The Hamiltonian given by Eq.\,(\ref{hsnss}) is diagonalised numerically and in our calculations  we consider realistic system parameters for InSb. 
$H_{{\rm S_{L}}{\rm N}}$ is the Hamiltonian that couples the  superconducting region ${\rm S}_{\rm{L}}$ to the normal region N, while $H_{{\rm N} {\rm S}_{\rm{R}}}$ couples N to ${\rm S}_{\rm{R}}$. 
The elements of these coupling matrices are non-zero only for adjacent sites that lie at the interfaces of the S and N regions. This is a good model when the barrier length is much shorter than the Fermi energy, otherwise see Ref.\,[\onlinecite{hoffman17}].
This barrier coupling is parametrized by a hopping $\uptau_{\rm L,R} v$, where $\uptau_{\rm L,R}\in[0,1]$ and $v$ is the spin-resolved hopping matrix in the wire. In general, we will asume a symmetric case where $\uptau_{\rm L,R}=\uptau$. 

In order to show that $\uptau$ indeed controls the transmission, we calculate the normal transmission 
$T_{\rm N}$ across the junction following the standard Green's functions technique\cite{PhysRevB.91.024514} assuming $\Delta=0$ and $\mu=0.5$\,meV. This is shown in Fig.\,\ref{Fig1}(c) for a short N region and different values of the Zeeman field. 
For $B=0$, $T_{\rm N}$ exhibits a non-linear dependence on $\uptau$ that quickly becomes zero for $\uptau\leq 0.6$, which signals the tunneling regime. 
%\comm{[This is tied to a specific choice of $\mu$ and $\Delta$. They should also be specified.] } 
The transparent regime is reached when $\uptau\approx 1$, as expected. Thus, the range $\uptau\in[0.6,1]$ allows us to explore the full range of transparencies from the tunnel to the transparent regime. By increasing $B$, the maximum transmission in the full transparent regime is halved, namely $T_{\rm N}=0.5$, due to the emergence of the helical state in the N region for $B>\mu$. Notice that the tunnel regime remains almost unaltered even when the Zeeman field is large. 
%The Hamiltonian given by Eq.\,(\ref{hsnss}) is diagonalised numerically; 

\subsection{Phase-dependent Andreev spectrum}
%Assuming decoupled superconducting regions sectors S$_{\rm L(R)}$ for the moment, 
At finite magnetic field the S regions develop two gaps $\Delta_{1,2}$ at low and high-momentum, respectively, and they exhibit a different behavior with Zeeman field $B$.
As $B$ increases beyond $B_{\rm c}$, each S region undergoes a topological phase transition, determined by the inversion of $\Delta_{1}$ at zero momentum, and enters into a topological superconducting phase.\cite{PhysRevLett.105.177002,Lutchyn:PRL10} The topological gap, sometimes called minigap, is then given by $\Delta_{2}$ for $B\gg B_{\rm c}$, which is controlled by the SO coupling. 
In this topological phase, MBSs emerge localized at each end of S$_{\rm L,R}$. Their wavefunctions decay in an exponentially and oscillatory fashion within a Majorana localization length $\xi_{\rm M}$ into the bulk of the superconducting regions,\cite{DasSarma:PRB12,Prada:PRB12,Klinovaja:PRB12} see Fig.\,\ref{Fig1}(b). As long as Majoranas do not overlap spatially, i.e. in long enough 
superconducting regions $L_{\rm S}\gg2\xi_{\rm M}$, they remain zero-energy eigenstates. A finite overlap, however, enables the hybridization of any two given Majoranas into a fermionic state of finite energy. When the left and right topological regions in the Josephson junction become coupled by a finite $T_{\rm N}$, the two inner Majoranas $\gamma_{2}$ and $\gamma_{3}$ also hybridise, although in a phase-dependent manner. This gives rise to a distinctive phase-dependent Andreev spectrum, as first discussed in Refs.\,\onlinecite{San-Jose:11a,Pikulin:PRB12,PhysRevB.91.024514,Hansen16}.

\begin{figure}
%\begin{minipage}[t]{\linewidth}
\centering
\includegraphics[width=.49\textwidth]{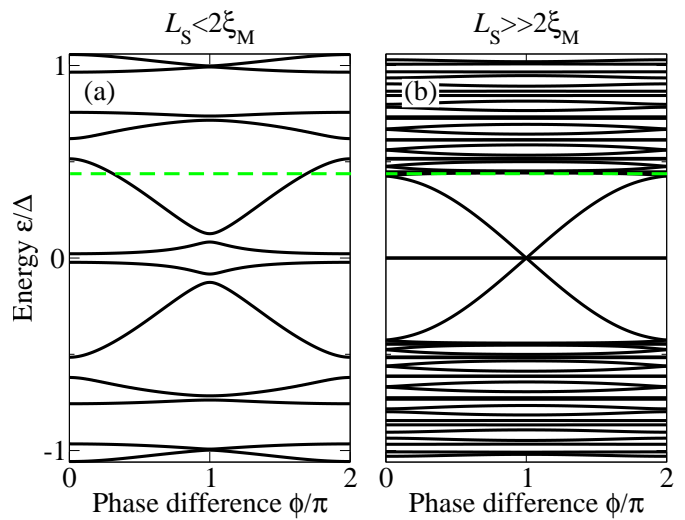} 
\caption{(Color online) Low-energy spectrum as a function of the superconducting phase difference in a short SNS junction for  $L_{\rm S}<2\xi_{\rm M}$ (a)  and 
$L_{\rm S}\gg2\xi_{\rm M}$ (b)  at $B=2B_{\rm c}$. The two lowest levels, almost insensitive to phase, correspond to the two outer Majoranas $\gamma_{1,4}$. The next two levels below the topological minigap (green dashed line), are strongly dispersive with phase, and come from the hybridization of the inner Majoranas $\gamma_{2,3}$ which anticross the former at $\phi=\pi$. Parameters: $L_{\rm N}=20$\,nm, 
(a) $L_{\rm S}=2000$\,nm, (b) $L_{\rm S}=10000$\,nm, $\alpha_{\rm R}=20$\,meVnm, $\mu=0.5$\,meV, $\Delta=0.25$meV.} 
\label{Fig2}
%\end{minipage}
\end{figure}

%Now we perform exact diagonalization of Eq.\,(\ref{hsnss}) and investigate the low-energy spectrum.
In Fig.\,\ref{Fig2} we present this low-energy spectrum %deep in the topological phase $B\gg B_{c}$ 
for $L_{\rm S}<2\xi_{\rm M}$ (a) and $L_{\rm S}\gg2\xi_{\rm M}$ (b). It consists of a discrete set of levels below the topological minigap (green dashed line), originated from the hybridization of the four Majoranas at the junction, and a quasi-continuum above. The lowest two levels are (almost) insensitive to $\phi$ and come from the outer MBSs, $\gamma_{1,4}$. The other two energy levels below the minigap originate from the inner Majoranas, $\gamma_{2,3}$, strongly disperse with $\phi$ and anticross at $\phi=\pi$ with the outer ones. For $L_{\rm S}<2\xi_{\rm M}$, Fig.\,\ref{Fig2}(a), a considerable splitting around $\phi=\pi$ appears, which arises due to the finite spatial overlap between the MBSs wave-functions within each S region. A true crossing is indeed reached when considering semi-infinite superconducting leads, $L_{S}\rightarrow\infty$. In such geometry the outer MBSs are not present in the description and the crossing at $\phi=\pi$ is protected by conservation of the total fermion parity. 

The splitting of the four MBSs at $\phi=\pi$ can be reduced  either by increasing the SO strength and therefore reducing the Majorana localization length,\cite{PhysRevB.91.024514} or by increasing the length of the superconducting regions $L_{\rm S}$, as shown in Fig.\,\ref{Fig2}(b). In the $L_{\rm S}\gg2\xi_{\rm M}$ limit, the outer MBSs are true Majorana zero modes, while the inner ones approach zero energy only at $\phi=\pi$. 
%Notice that in this regime the four lowest levels are truly located within the minigap. 
As we argue in what follows, the critical current across the junction is very sensitive to the $\phi=\pi$ anticrossing structure and can, therefore, be used to extract useful information about the Majorana hybridizations in the system.

\begin{figure}
\begin{minipage}[t]{\linewidth}
\centering
\includegraphics[width=.99\textwidth]{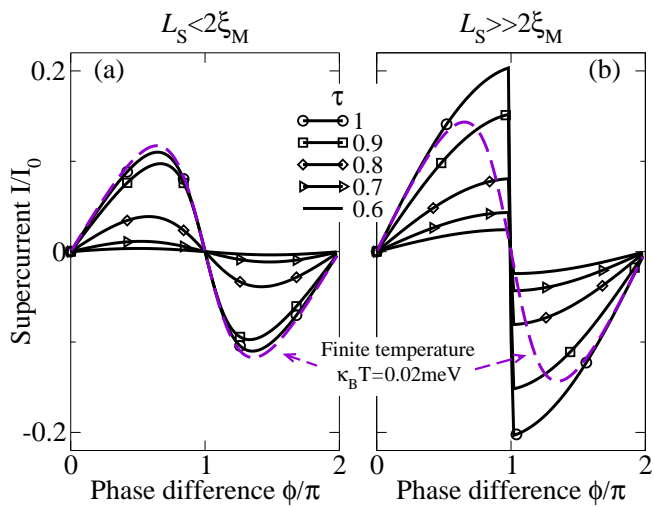} \\
\caption{(Color online) Supercurrent in a short SNS junction as a function of $\phi$ at $B=2B_{\rm c}$ for  $L_{\rm S}< 2\xi_{\rm M}$ (a) and $L_{\rm S}\gg 2\xi_{\rm M}$ (b). Different curves correspond to different values of $\uptau$. A sawtooth profile in (b) is developed when MBS wavefunctions within each superconducting section do not overlap, Fig.\,\ref{Fig1}(b).
This feature remains robust as $\uptau$ is reduced, but not so with finite temperature (see dashed purple curve, with $\kappa_{B}T=0.02$\,meV and $\uptau=1$).  Parameters: $L_{\rm N}=20$\,nm, (a) $L_{\rm S}=2000$\,nm, (b) $L_{\rm S}=10000$\,nm, $\alpha_{\rm R}=20$\,meVnm, $\mu=0.5$\,meV, $\Delta=0.25$meV and $I_{0}=e\Delta/\hbar$.}
\label{Fig3}
\end{minipage}
\end{figure}

\subsection{Current-phase characteristics}
The supercurrent can be obtained from the free energy of the junction as
\begin{equation}
I(\phi)=\frac{1}{\Phi_0}\frac{dF}{d\varphi}, 
\label{free-energy}
\end{equation}
where $\Phi_0=\frac{\hbar}{2e}$ is the reduced (superconducting) flux quantum. Taking care about the double counting inherent to the BdG description, the above expression can be rewritten as
\begin{equation}
\label{shortJosephcurrent2}
%I(\phi)=-\frac{e}{\hbar}\sum_{p>0}\frac{d\varepsilon_{p}}{d\phi}\,.
I(\phi)=-\frac{e}{\hbar}\sum_{\epsilon_p>0}{\rm tanh}\Big(\frac{\varepsilon_{p}}{2\kappa_{B}^{}T} \Big)\frac{d\varepsilon_{p}}{d\varphi}\,,
\end{equation}
where $\kappa_{\rm B}$ is the Boltzman constant, $T$ is the temperature and $\varepsilon_{p}$ are the energy levels, obtained by exact diagonalization of the Hamiltonian in Eq. \eqref{hsnss}. Importantly, the sum must be performed over all degeneracies (spin in our case) explicitly.\footnote{Note the missing factor 2 as compared to, e. g., Eq. (2.16) in Ref. [\onlinecite{Beenakker:92}]. There, spin degeneracy is assumed, hence the degeneracy factor 2 [this should not be confused with the factor 2 of Eq. (\ref{free-energy}) contained in $\Phi_0$, which comes from the charge $2e$ of Cooper pairs]. In our case, spin degeneracy is broken by the Zeeman field $B$, so the spin summation must be done explicitly, and no degeneracy factor 2 should be included.
% such that the expression is valid both for spinful and spinless cases as the Zeeman field increases.
} In what follows, we focus on zero temperature calculations, unless otherwise stated, and short junctions. The full phenomenology of current-phase relationships, including the long junction limit, will be discussed elsewhere.\cite{cayao17}
%Eq.\,(\ref{shortJosephcurrent2}) allows us to include the discrete quasi-continuum contribution in the calculation of the  supercurrent $I(\phi)$. Although, Eq.\,(\ref{shortJosephcurrent2}) allows us for calculating the supercurrent in both short and long SNS junctions, we only discuss short junctions.

Fig.\,\ref{Fig3} shows the phase-dependent supercurrents corresponding to the Andreev spectra of Fig.\,\ref{Fig2}. The finite splitting at $\phi=\pi$, observed for $L_{\rm S}<2\xi_{\rm M}$, gives rise 
to supercurrents with a ${\rm sine}$-like behavior as a function of $\phi$, see Fig.\,\ref{Fig3}(a). The opposite case of $L_{\rm S}\gg2\xi_{\rm M}$ that corresponds to an almost perfect crossing at $\phi=\pi$ results in a sawtooth profile, Fig.\,\ref{Fig3}(b). Note how the overall shape in both limits remains unaltered irrespective of the transmission across the normal region all the way from fully transparent junctions $\uptau=1$ to the tunnel limit 
$\uptau=0.6$. 
As expected, finite temperature (dashed magenta curves), however, washes out the sawtooth profile. %\textcolor{red}
\comm{Thus, current-phase curves cannot be used to clearly distinguish the trivial and topological phases based on the sawtooth profile at $\phi=\pi$. Despite this, the critical current still contains useful information in the topological phase, as we discuss now.}

\section{Critical currents}
\label{sec13}
The critical current $I_{\rm c}$ is the maximum supercurrent that can flow through the junction. We calculate it by maximizing the  supercurrent $I(\phi)$ with respect to the superconducting 
phase difference $\phi$, 
\begin{equation}
I_{\rm c}={\rm max_{\phi}}[I(\phi)]\,.
\end{equation}
%We stress that previous equation remains valid for any length of the normal region, although the calculation 
%requires long calculation times. As discussed before, in this work we only focus on short SNS junctions.
%Before going further we firstly discuss the validity of previous equation in short and long junctions. 

\begin{figure}
\begin{minipage}[t]{\linewidth}
\centering
\includegraphics[width=.99\textwidth]{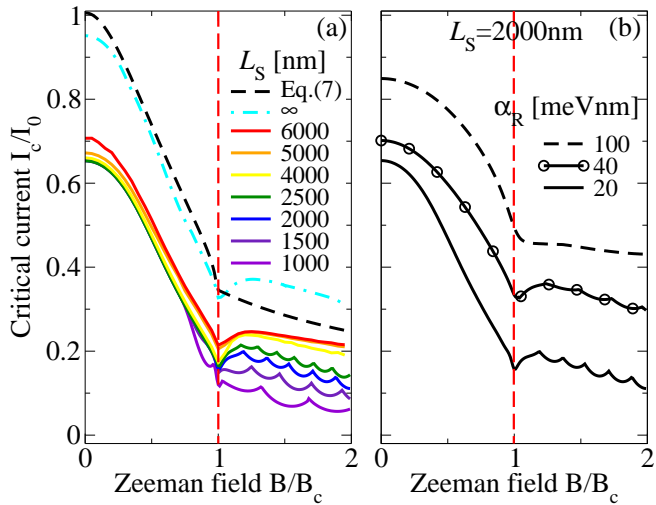} \\
\caption[Critical current in a short SNS junction]{(Color online) Critical current $I_{\rm c}$ in a short and transparent SNS junction as a function of the Zeeman field $B$. In (a) different curves represent different values of the superconducting region length $L_{\rm S}$. As $B$ increases towards $B_{\rm c}$,  $I_{c}$ decreases, approximately following the evolution of the gap. At the gap closing $B=B_{\rm c}$ (red vertical dashed line),  $I_{\rm c}$ exhibits a non-trivial feature but remains finite. For $B>B_{\rm c}$, the critical current develops an oscillatory behavior that is related to the finite inner-outer Majorana splitting and disappear as the length $L_{\rm S}$ is increased. Notice that the magnitude of $I_{\rm c}$ for $B>B_{\rm c}$ \emph{increases} with increasing $L_{\rm S}$ while its value for $B<B_{\rm c}$ remains unchanged. This unconventional dependence is a direct signature of nontrivial topology of the junction.
(b) $I_{c}$ for fixed $L_{\rm S}=2000$\,nm and increasing  SO strength $\alpha_{\rm R}$.
Observe that at large fields and in the very strong SO regime the critical current saturates to $1/2$ of its $B=0$ value. 
Parameters: $L_{\rm N}=20$\,nm, $\alpha_{\rm R}=20$\,meVnm, $\mu=0.5$\,meV, $\Delta=0.25$meV and $I_{0}=e\Delta/\hbar$.
}
\label{Fig4}
\end{minipage} 
\end{figure}

In Fig.\,\ref{Fig4} we show $I_{\rm c}$ for short and transparent SNS 
junctions as the Zeeman field $B$ increases from the trivial ($B<B_{\rm c}$) to the 
topological phase ($B>B_{\rm c}$).
Different curves in Fig.\,\ref{Fig4}(a) correspond to different values of the superconducting region length $L_{\rm S}$. The critical current is maximum at zero field and decreases as the Zeeman field $B$ increases, tracing the reduction of the low-momentum gap $\Delta_{1}$.

\begin{figure}
\begin{minipage}[t]{\linewidth}
\centering
\includegraphics[width=.99\textwidth]{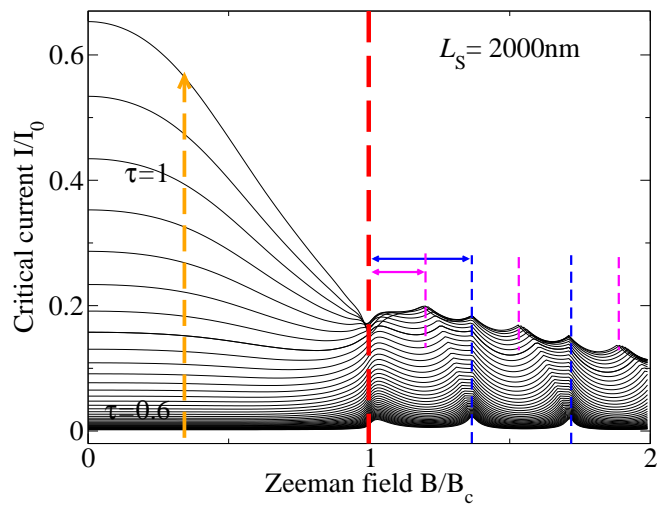} \\
\caption[Critical current in a short SNS junction]{(Color online) $I_{\rm c}$ for fixed $L_{\rm S}=2000$\,nm and increasing values of $\uptau\in[0.6,1]$, in steps of $0.01$ (indicated by vertical orange arrow). In the trivial phase, $B<B_{\rm c}$, the critical current is quickly suppressed as $\uptau$ is decreased towards the tunneling limit. In the topological phase the critical current is suppressed much more slowly with $\uptau$, which results in a re-entrant $I_{\rm c}$ at $B_{c}$ for small $\uptau$. For $B>B_{\rm c}$, the period of the $I_{\rm c}$ oscillations is doubled as the transparency of the junction is reduced (magenta and blue dashed lines). 
Parameters are the same as in Fig.\,\ref{Fig3}.
%$L_{\rm N}=20$\,nm, $\alpha_{\rm R}=20$\,meVnm, $\mu=0.5$\,meV, $\Delta=0.25$meV and $I_{0}=e\Delta/\hbar$.
}
\label{Fig5}
\end{minipage} 
\end{figure}

Remarkably, the energy gap closes at $B=B_{\rm c}$ but the critical current remains finite and develops a robust kink-like feature, {as can be seen in Fig.\,\ref{Fig4}(a)}. At $B=B_{\rm c}$, $I_{\rm c}$ is roughly halved with respect to its $B=0$ value, as a result of the system losing one channel  (the nanowire bands change from spinful to spinless). As the Zeeman field is increased further, the critical current develops oscillations when $L_{\rm S}\lesssim 2\xi_{\rm M}$. This structure arises from the anticrossing of the four MBSs around $\phi=\pi$. The oscillations disappear  as the length $L_{\rm S}$ is increased, and the outer Majoranas decouple from the junction, {as can be seen in Fig.\,\ref{Fig4}(a)}. Another interesting feature is that for $B<B_{\rm c}$, $I_c$ remains almost unaltered with $L_{\rm S}$, as expected of a conventional ballistic junction. However, for $B>B_c$, $I_c$ \emph{increases} as $L_{\rm S}$ becomes longer, an effect that can be traced back to the protected crossing at $\phi=\pi$ of the topological Andreev spectrum, see Fig. \ref{Fig2}(b). Ultimately, % \textcolor{red}{and quite remarkably}, 
the  increase in critical current is therefore related to the Majorana origin of the low-energy Andreev bound states in the junction. This remarkable length-dependence in a short and transparent junction when $B>B_{\rm c}$ is thus a direct signature of the nontrivial topology and provides a way to distinguish the effects described here from other trivial explanations such as $\pi$-junction behavior or oscillatory critical currents due to orbital effects.\cite{zuo17} 

In the asymptotic limit $L_{\rm S}\rightarrow\infty$ the outer MBSs are not involved in the supercurrent and therefore only the inner ones determine the critical current. This limit is shown by the dot-dashed curve in Fig.\,\ref{Fig4}(a), which matches quite accurately a simplified model for $I_c$ coming purely from the inner Majoranas, see black dashed line. This model, valid in the transparent limit, yields\cite{SanJoseNJP:13}
\begin{equation}
\label{Eq7}
\begin{split}
I_{\rm c}&\approx\frac{I_0}{2}\frac{\Delta_1+\Delta_2}{\Delta}, \,\, B<B_{\rm c} \\
I_{\rm c}&\approx\frac{I_0}{2}\frac{\Delta_2}{\Delta}, \,\, B>B_{\rm c},
\end{split}
\end{equation}
with $\Delta_1=|B-B_{\rm c}|$ and 
\begin{equation}
\Delta_2\approx\frac{2\Delta E_{\rm SO}}{\sqrt{E_{\rm SO}(2E_{\rm SO}+\sqrt{B^2+4E^2_{\rm SO}})}}\nonumber,
\end{equation}
the low and high-momentum gaps introduced before, respectively. Here, $I_0=e\Delta/\hbar$ is the maximum critical current supported by an open channel. The critical current in this simple model results from the contribution of two Andreev bound states within $\Delta_{1}$ and $\Delta_{2}$ for $B<B_{\rm c}$, and only one bound state within $\Delta_{2}$ for $B>B_{\rm c}$. As in the exact numerical calculation, the ratio $I_{\rm c}/I_0$ decreases as the Zeeman field increases, which reflects the transition from spinful to spinless bands at $B=B_c$.\cite{SanJoseNJP:13} 
 
Fig.\,\ref{Fig4}(b) shows the dependence of $I_{\rm c}(B)$ with SO coupling. In the strong SO limit (dashed curve), the outer gap becomes $\Delta_2\approx\Delta$ and thus the ratio between the critical currents at zero and high Zeeman fields reaches 1/2, which is yet another strong signature of a topological phase. Note, however, that the required SO coupling to reach this ratio may be unrealistically large in most nanowire systems.

The preceding results correspond to transparent junctions. In Fig.\,\ref{Fig5} we show the critical current as a function of $B$ for different values of $\uptau$, or junction transparency. As expected, $I_c$ in the trivial regime $B<B_{\rm c}$ gets reduced as $\uptau$ decreases, until it becomes negligible in the tunneling regime. The kink-like feature in $I_{c}$ for a transparent $\uptau\approx 1$ junction associated to the closing of the gap at $B=B_{\rm c}$ evolves into a step-like increase as $\uptau$ is reduced. 
This causes a re-entrant critical current in the tunneling regime, where $I_c$ is essentially zero below $B_{\rm c}$ but then increases suddenly at the topological phase transition. The re-entrance of the low-transparency $I_c$ as $B$ exceeds $B_{\rm c}$ can be understood from the fact that the critical current carried by the inner Majoranas scales as $I_{\rm c}\sim\sqrt{T_{\rm N}}$  for $B>B_{\rm c}$. 
At small $T_{\rm N}$ this current is much larger than the critical current $I_{\rm c}\sim T_{\rm N}$ associated to conventional Andreev levels for $B<B_{\rm c}$.\cite{PhysRevLett.112.137001,tiira17}  Importantly, we also find that the critical current oscillations for $B>B_{c}$ show a characteristic period doubling as $\uptau$ is reduced, see Fig.\,\ref{Fig5}. This period doubling is highlighted by the magenta and blue vertical dashed lines.

\begin{figure}
\begin{minipage}[t]{\linewidth}
\centering
\includegraphics[width=.99\textwidth]{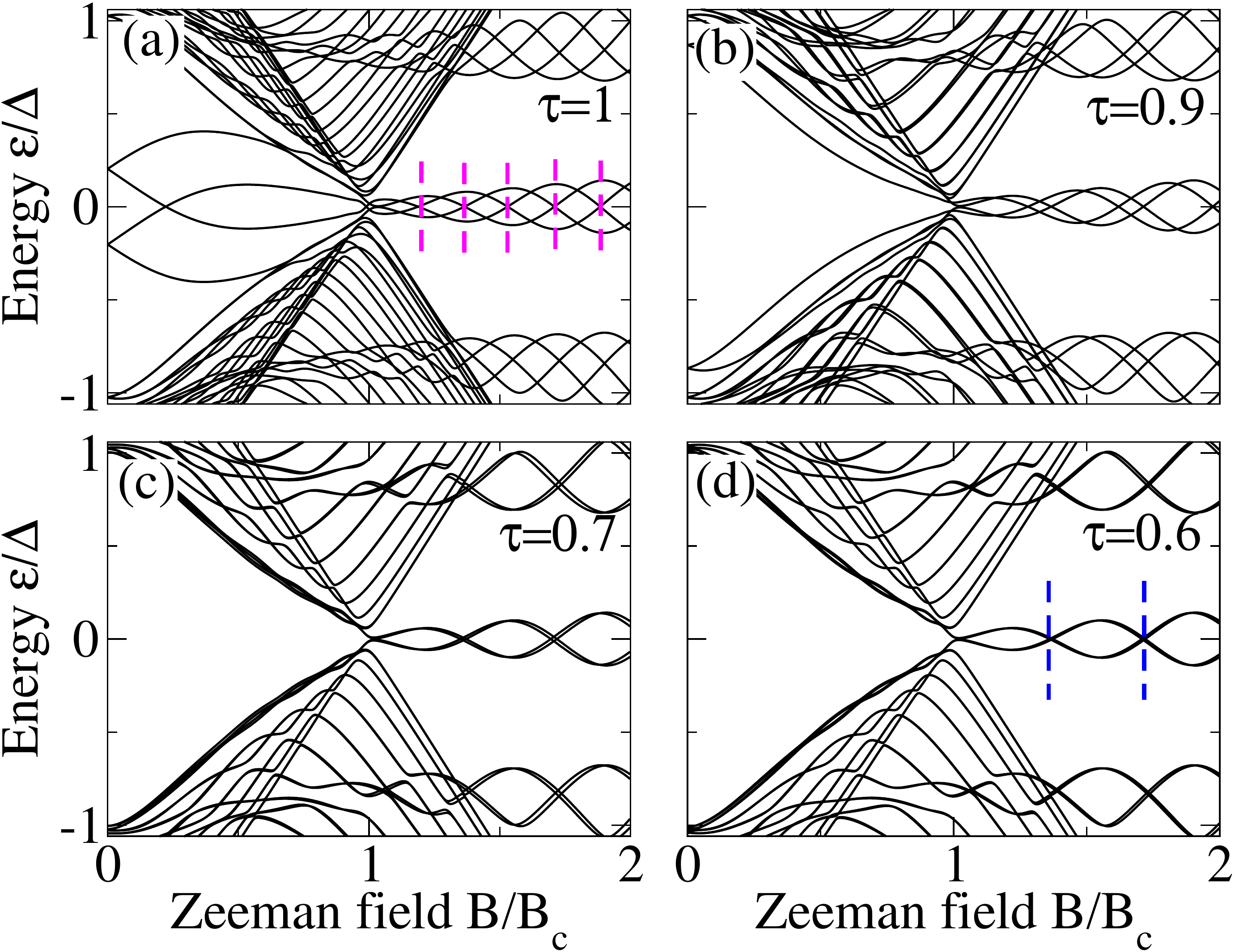} 
\caption{(Color online) Low-energy Andreev spectrum as a function of the Zeeman field in a short SNS junction at $\phi=\pi$ for different values of the coupling between N and S regions 
$\uptau$. Note that upon decreasing $\uptau$ the low-energy zero-energy crossings for $B>B_{\rm c}$ merge in pairs, which results in a doubling of the period of oscillation in $I_c$. Parameters same as in Fig.\,\ref{Fig3} and $L_{\rm S}=2000$\,nm.
%Parameters: $L_{\rm N}=20$\,nm, $L_{\rm S}=2000$\,nm, $\alpha_{\rm R}=20$\,meVnm, $\mu=0.5$\,meV and $\Delta=0.25$\,meV.
}
\label{Fig6}
\end{minipage}
\end{figure}

\begin{figure}
\begin{minipage}[t]{\linewidth}
\centering
\includegraphics[width=.99\textwidth]{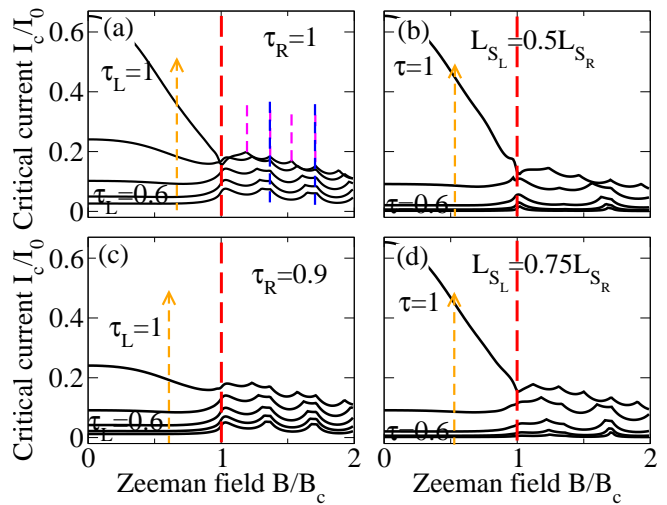} 
\caption{(Color online) Critical currents as a function of the Zeeman field in a short SNS junction. (a,c) Asymmetry in the couplings to the left and right superconducting leads, $\uptau_{\rm L(R)}$, for fixed $L_{\rm S}$. The coupling to the right S region $\uptau_{\rm R}$ is fixed and the one for the left one varies from tunnel to the full transparent regime in steps of $0.1$. This situation is expected to arise in experiments. Observe that the oscillations for $B>B_{\rm c}$ remain robust  with a clear period doubling for decreasing $\uptau_{\rm L}$.
(b,d) Asymmetry in the length of the superconducting regions $L_{\rm S}=2000$\,nm for different values of the coupling parameter $\uptau_{\rm L(R)}=\uptau$ from tunnel to the full transparent regime in steps of $0.1$.
Parameters: $L_{\rm N}=20$\,nm, $\alpha_{\rm R}=20$\,meVnm, $\mu=0.5$\,meV and $\Delta=0.25$\,meV.
}
\label{Fig7}
\end{minipage}
\end{figure}

\begin{figure}
\begin{minipage}[t]{\linewidth}
\centering
\includegraphics[width=.99\textwidth]{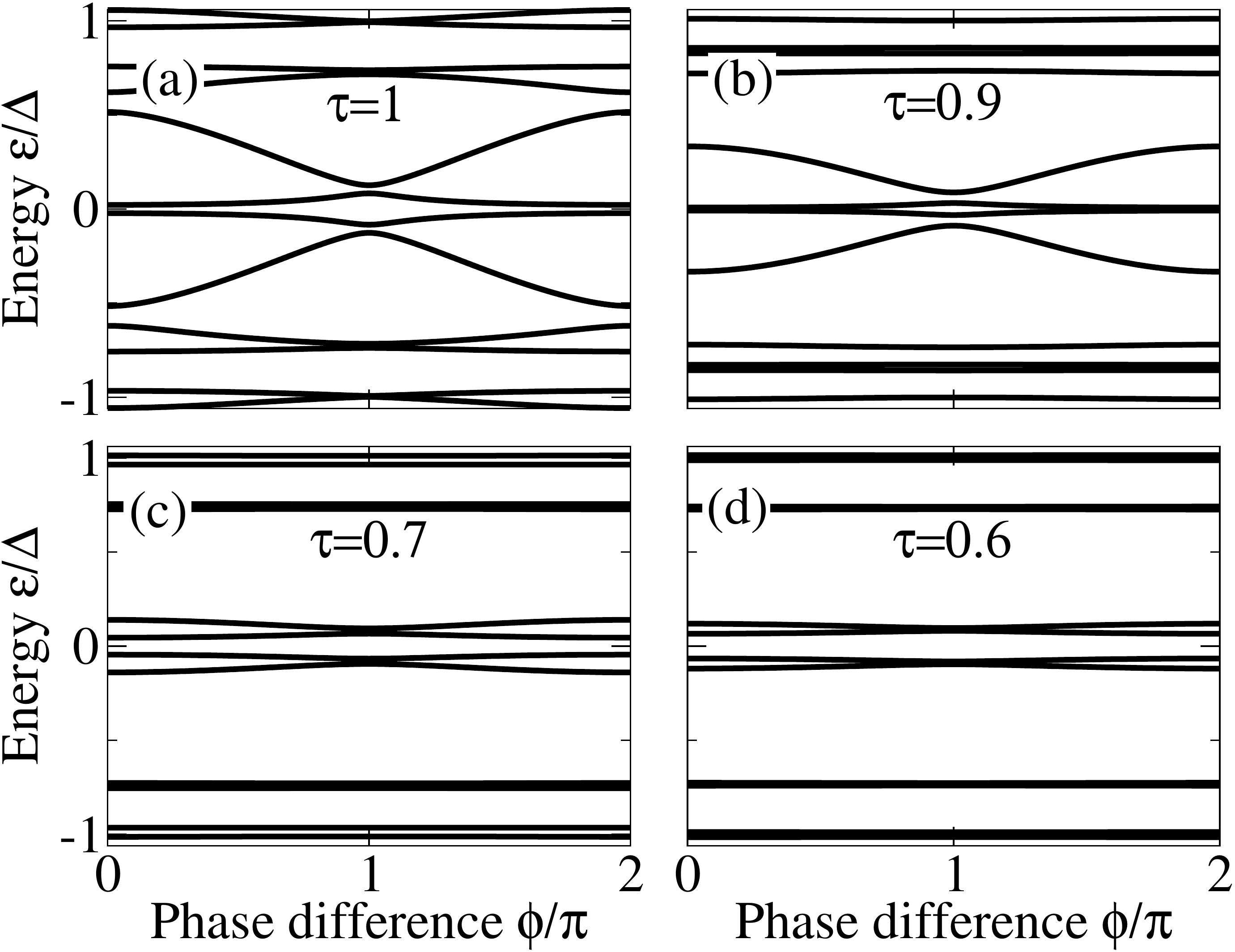} 
\caption{(Color online) Low-energy Andreev spectrum as a function of the superconducting phase difference $\phi$ at $B=2B_{c}$ for different values of the coupling between N and S regions $\uptau$. Note that upon decreasing $\uptau$ the outer (two lowest) MBSs acquire a finite energy at $\phi=2\pi n$, for $n=0,1,2,\dots$, while developing a crossing at $\phi=\pi$ with their inner counterparts. This is the limit of two identical decoupled superconducting nanowires with overlapping Majoranas at each end. Parameters same as in Fig.\,\ref{Fig3} and $L_{\rm S}=2000$\,nm.
%Parameters: $L_{\rm N}=20$\,nm, $L_{\rm S}=2000$\,nm, $\alpha_{\rm R}=20$\,meVnm, $\mu=0.5$\,meV and $\Delta=0.25$\,meV.
}
\label{Fig8}
\end{minipage}
\end{figure}

The period doubling in the topological regime $B<B_{\rm c}$  can be understood in terms of changes in the structure of the $\phi=\pi$ anticrossing of the four hybridised MBSs in the system as $\uptau$ is reduced is presented in Fig.\,\ref{Fig6}. Around this point the four states are close to zero and have a maximal slope with $\phi$. Hence the supercurrent $I(\phi)$ from the MBSs reaches its maximum value close to $\phi=\pi$ as we saw in Figs.\,\ref{Fig2} and \ref{Fig3}. The structure of the low-energy $\phi=\pi$  anticrossing therefore controls $I_{\rm c}$. The anticrossing results from the $B$-dependent hybridization of the four MBSs, and exhibits an oscillatory pattern with $B$. The oscillatory structure of the resulting $\phi=\pi$ spectrum in a fully transparent junction ($\uptau=1$) is shown in Fig.\,\ref{Fig6}(a).
As the coupling $\uptau$ is reduced, Fig.\,\ref{Fig6}(b-c), two consecutive zero-energy crossings (magenta vertical dashed lines in panel a) tend to approach each other and finally fuse into one single crossing in the tunneling regime (blue vertical dashed lines in Fig.\,\ref{Fig6}(d)). In this limit, the periodicity is that of overlapping Majoranas in essentially decoupled wires, namely two independent pairs of Majoranas which do not couple through the normal region. 

{We now analyze the robustness of the period-doubling effect against various asymmetries of the system. In Fig. \ref{Fig7} we plot $I_{\rm c}$ as a function of $B$ for asymmetries in (a,c) the parameter $\uptau_{\rm L,R}$ and (b,d) the length of the S regions. Remarkably, by fixing $\uptau_{\rm R}$ and varying $\uptau_{\rm L}$ the period doubling remains robust indicating that this effect may be observable in real experiments, as asymmetric junctions is the natural experimental situation. Although the asymmetry of the length of the S regions suppresses the visibility of the period doubling, the effect is still appreciable.}

The period doubling effect can be also seen in the phase-dependent Andreev spectra. This is shown in Fig. \ref{Fig8}, {where different panels correspond to different values of $\uptau$. Observe that as $\uptau$ is reduced the inner and the outer MBSs are strongly affected. In fact, the inner MBSs reduce their energies at $\phi=0$ but the outer ones slightly increase, while keeping almost a constant value around $\phi=\pi$. In the tunneling regime, one outer and one inner MBS, which correspond to the same S region,  get paired with a small dependence on $\phi$.}
 Therefore, the $I_{\rm c}$ oscillations and the period doubling upon reducing the transparency of the junction (with e.g. a central gate) can be understood in terms of Majorana hybridisations in the system. 
%This, together with the aforementioned anomalous $L_S$ dependence for $B>B_{c}$, provides an alternative detection scheme of Majoranas and their energy splittings beyond standard tunnel spectroscopy experiments.

\section{Conclusions}
\label{concl}
We have theoretically investigated how Majorana physics manifests in Josephson junctions based on Rashba nanowires proximitized with superconductors. 
We have demonstrated that the critical current in short junctions between two finite length superconductors and tunable normal transmission $T_{\rm N}$ 
contains important information about nontrivial topology and MBSs.
For $B<B_{\rm c}$, $I_c$ does not depend on the superconducting region length, as expected for a trivial ballistic junction in the short junction regime. In contrast, the critical current in the topological regime increases with $L_{\rm S}$. This anomalous length-dependence in a short SNS junction is a direct consequence of Majorana overlap on each S region of the junction.
We have observed a re-entrant effect in $I_c$ in the tunneling regime which results from an enhanced Majorana-mediated critical current.
$I_c$ is essentially zero below $B_{\rm c}$ but increases abruptly at the topological phase transition.
Importantly, we have also found that the critical current for $B>B_{\rm c}$ exhibits oscillations derived from the oscillatory splitting of overlapping Majoranas around zero energy, which are robust against temperature and changes in $T_{\rm N}$. Tuning the latter, in particular, gives rise to a period doubling effect in the oscillations  which can be unmistakably attributed to overlapping MBSs. 
%\textcolor{red}

\comm{Interestingly, these findings are robust under finite temperatures (below the induced gap), as well as under finite disorder.} {Our findings  complement previous studies\cite{PhysRevLett.112.137001, tiira17} in SNS junctions and, quite remarkably, based on a recent experiment,  we believe that our proposal is  experimentally reachable.\cite{tiira17}} 
%, thus implying the experimental feasibility of our findings.}
%We have also shown that the expected sawtooth profile in the phase-dependent supercurrent for $L_{\rm S}\gg 2\xi_{\rm M}$ {when MBSs at each S region do not overlap}, resulting from almost perfect crossings at $\phi=\pi$, does not depend on the transparency of the normal section. On the contrary, the sawtooth shape is rapidly lost as the finite length of the superconducting sections becomes relevant, namely when $L_{\rm S}\leq2\xi_{\rm M}$. In this limit, sizable energy splittings around $\phi=\pi$ resulting from Majorana overlaps, give rise to increasingly conventional current-phase characteristics. 
%This makes it hard to distinguish trivial from non-trivial topology from current-phase measurements, unlike critical currents, in short wires. \textcolor{red}{\textbf{[This last bit is dangerous overselling, in my opinion (PSJ). $I(\phi)$ always contains more information than $I_c$.]}}
Moreover, all these features can be modified by tuning the normal transmission across the junction, thus providing further evidence of Majoranas and their energy splittings beyond tunnel spectroscopy.
%Despite this, we demonstrate that the critical current still contains important information about nontrivial topology and Majoranas. This includes signatures of the gap inversion at the topological transition, a re-entrant effect,\cite{PhysRevLett.112.137001,tiira17} an oscillatory pattern that originates from Majorana overlaps, a period-doubling effect as a function of junction transparency, and an anomalous length dependence that can be also traced back to the presence of Majoranas in the junction. All these features can be modified by tuning the normal transmission across the junction, thus providing an alternative detection scheme of Majoranas and their energy splittings beyond tunnel spectroscopy.

\section*{Acknowledgements}
We acknowledge financial support from the Spanish Ministry of Economy and Competitiveness through Grant No. FIS2015-65706-P (MINECO/FEDER) (P. S.-J), FIS2015-64654-P  (R. A.), FIS2016-80434-P (AEI/FEDER, EU) (E. P.) and the Ram\'{o}n y Cajal programme through grant No. RYC-2011-09345 (E. P). 
J.C. and A.B.S. acknowledge financial support from the Swedish Research Council (Vetenskapsr\aa det), the G\"{o}ran Gustafsson Foundation, the Swedish Foundation for Strategic Research (SSF), and the Knut and Alice Wallenberg Foundation through the Wallenberg Academy Fellows program.
%\clearpage

%\onecolumngrid
%\clearpage
%\twocolumngrid

% -------------------------------------- %
% APPENDIX:
% -------------------------------------- %
\appendix

\section{Finite temperature, disorder and Majorana localization length}
\label{Appendix}
We have shown in the main text that at zero temperature the critical current exhibits a reentrant behavior in the tunneling regime at $B=B_{\rm c}$, oscillations in the topological phase $B>B_{\rm c}$ and a unique period doubling effect. In this part, we present additional calculations that support the robustness of these findings.

\begin{figure}[b]
%\begin{minipage}[t]{\linewidth}
\centering
\includegraphics[width=.49\textwidth]{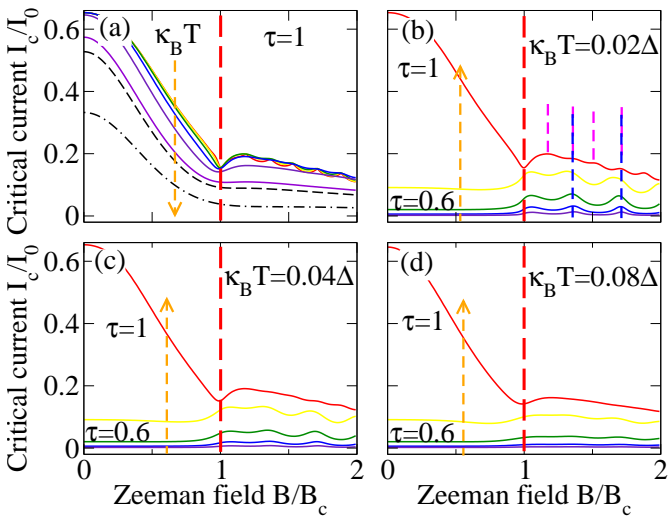} 
\caption{(Color online) Effect of finite temperature on the critical currents as a function of the Zeeman field in a short SNS junction. 
(a) Full transparent junction $\uptau=1$ and increasing temperature (top-to-bottom curves). From top to bottom: red, yellow, green, blue, indigo, violet, dashed black and dash-dot black curves correspond to $\kappa_{\rm B}T=(0,0.01,0.02,0.04,0.08,0.16,0.2,0.4)\Delta$. (b-d) Different panels correspond to a different temperature value for different cases of $\uptau$. Observe that the reentrant behavior, oscillations for $B>B_{\rm c}$ and period doubling effect remain robust at finite temperatures, but much less than the superconducting gap.
Parameters: $L_{\rm N}=20$\,nm, $\alpha_{\rm R}=20$\,meVnm, $\mu=0.5$\,meV and $\Delta=0.25$\,meV.
}
\label{FigApp1}
%\end{minipage}
\end{figure}

First, we discuss the effect of finite temperature on the critical current $I_{\rm c}$, where we employ Eq.\,(\ref{shortJosephcurrent2}). The results are presented in Fig.\,\ref{FigApp1}. In Fig.\,\ref{FigApp1}(a) we show the critical current for different temperature values in a transparent junction $\uptau=1$. In Fig.\,\ref{FigApp1}(b-d) different panels correspond to different values of temperature for different values of the normal transmission $\uptau$.
We observe that at finite, but small, temperature the critical current remains robust. In particular, one notices that the reentrant behavior at $B=B_{\rm c}$ of the critical current in the tunneling regime is visible, as well as the oscillations in the topological phase $B>B_{c}$. Remarkably, the period doubling effect of the critical current oscillations in the topological phase also remain visible at finite temperature. 
As expected, high temperature values tend to wash out the oscillations in the critical current, as one can indeed see in the three lowest curves of Fig.\,\ref{FigApp1}(a). %This is indeed an effect that the majority of observables suffer in condensed matter, where really low temperatures are necessary.

\begin{figure}[b]
%\begin{minipage}[t]{\linewidth}
\centering
\includegraphics[width=.49\textwidth]{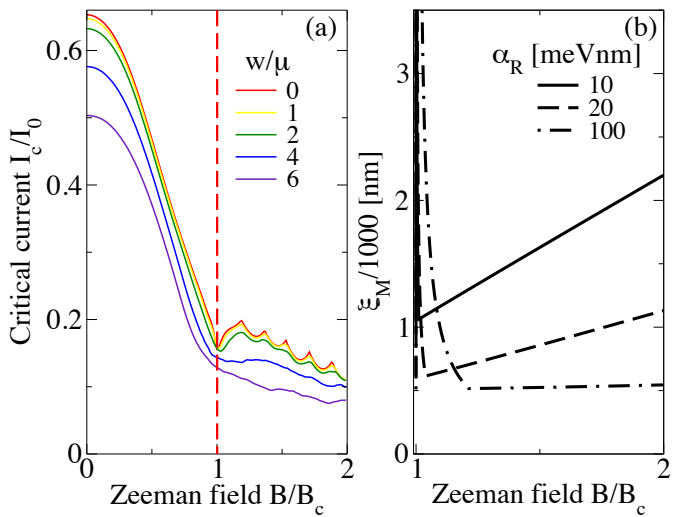} 
\caption{(Color online) (a) Effect of disorder on  the critical currents as a function of the Zeeman field in a short SNS junction. Notice that the oscillations for $B>B_{\rm c}$ persist even when the strength of disorder is comparable to the value of the chemical potential $\mu$. 
(b) Majorana localization length as a function of $B$ for different values of the SOC strength.
 Parameters: $L_{\rm N}=20$\,nm, $\alpha_{\rm R}=20$\,meVnm, $\mu=0.5$\,meV and $\Delta=0.25$\,meV.
}
\label{FigApp2}
%\end{minipage}
\end{figure}

Another important effect that we discuss here is the role of on-site disorder, which  introduces random fluctuations in the chemical potential of the system $\mu$. It is introduced as a random onsite potential $V_{i}$ in the tight-binding form of Hamiltonian given by Eq.\,(\ref{H0Hamil}), where  the values of $V_{i}$ lie within $[-w,w]$, being $w$ the disorder strength.
Fig.\,\ref{FigApp2}(a) presents the critical current as a function of the Zeeman field for different disorder strengths, where each curve corresponds to 20 disorder realizations. A remarkable feature here is that the oscillations of the critical current in the topological phase persist even for $w$ slightly stronger than the chemical potential $\mu$ (red, yellow and green curves). For very strong disorder, however, the visibility of the oscillations is considerable degraded (blue and indigo curves). 
This, together with the discussion about finite temperature, therefore  suggests that our proposal represents an experimental feasible route for the search of MBSs.

For completeness we also show the dependence of the Majorana localization length $\xi_{\rm M}$ on the Zeeman field for different values of the SOC strength. It is obtained after solving numerically the polynomial equation\cite{Lutchyn:PRL10,PhysRevB.91.024514}  $k^{2}+4(\mu+C\alpha_{R}^{2})Ck^{2}+8\lambda C^{2}\Delta \alpha_{R} k+4C_{0}C^{2}=0\,,$ where $C=m/\hbar^{2}$ and $C_{0}=\mu^{2}+\Delta^{2}-B^{2}$. This equation is obtained from Eq.\,(\ref{H0Hamil}) with induced superconducting potential $\Delta$. We define the Majorana localization length as $\xi_{\rm M}={\rm max}[-1/k_{sol}]$, where $\{k_{sol}\}$ are solutions to the previous polynomial equation. Observe that for realistic parameters (dashed line) $\xi_{\rm M}$ is large implying that one needs very long wires to avoid the energy splitting of MBSs. These results are indeed in good agreement with the conditions for the ratio between the superconducting region lengths  and Majorana localization length presented in the main text when discussing Majorana splitting in the SNS junction setup.

\bibliography{biblio}

\end{document}